\documentstyle[11pt,newpasp,twoside,epsf]{article}
\def\edcomment#1{\iffalse\marginpar{\raggedright\sl#1\/}\else\relax\fi}
\reversemarginpar
\begin{document}
\title{Application of DSMC method to astrophysical flows}

\author{Takuya Matsuda \& Hiromi Mizutani}
\affil{Dept. of Earth and Planetary Sciences, Kobe University, Kobe 657-8501, Japan}
\author{Henri.M.J.Boffin}
\affil{Royal Observatory of Belgium, 3 av. Circulaire, 1180 Brussels}
\begin{abstract}
The Direct Simulation Monte Carlo (DSMC) method, developed originally to calculate
rarefied gas dynamical problems, is applied to continuous flow including
shocks assuming that the Knudsen number is sufficiently small. In particular, we study the formation of spiral shocks in the accretion disc of a close binary system. The method involves viscosity and thermal conduction automatically, and can thus simulate turbulent viscosity. 
\end{abstract}
\section{The Direct Simulation Monte Carlo Method}
The Direct Simulation Monte Carlo method was invented by Bird (see Bird 1994). 
It was originally developed to treat rarefied
gas flows. In DSMC, the microscopic molecular motion and their mutual collisions are
calculated. Macroscopic quantities such as density, velocity and temperature
of fluids are calculated by taking some mean of microscopic quantities. Molecular collisions are handled in a stochastic manner. The collision model in
DSMC is such that the molecules, or more precisely, the collection of molecules considered as a particle, are assumed to be hard billiard balls.
 
We divide space into many cells, each
containing many such model molecules. Particles in a cell are assumed to have
the possibility of mutual collisions.  The direction of
the particle motion after the collision is decided stochastically.
Density can then be
calculated by using the particles number in a cell. The fluid velocity in a cell is nothing but a mean velocity of
molecules in the cell and we obtain the thermal velocity of a particle by
subtracting the mean velocity in the cell from the particle velocity. We may
then define the temperature of the gas using the thermal velocity of molecules.

In the process of collision, the number and the total momentum of two
particles are conserved. Total kinetic energy of two particles is conserved in an
adiabatic case, but it may not conserve if we include cooling. We consider
two method of cooling (see below). 

\section{Merits of DSMC}
As was stated earlier, DSMC has been developed to treat rarefied gas flow. In
this sense the method is a scheme to solve the Boltzmann equation. If the mean
free path of the particles $\lambda$ is much less than the typical length scale,
$L$, we may derive the Navier-Stokes equations from the Boltzmann equations.
Defining the Knudsen number as $K_n=\lambda/L$, we see that in the limit of small
$K_n$, DSMC can handle continuum flows: $K_n=0.01-0.1$ is good enough for that
purpose.
 Since we solve the molecular motion, molecular viscosity and molecular thermal
conduction are included automatically in the system.  In the kinetic theory of
gas, we know that the dynamical viscosity coefficient $\nu$ is estimated to be
$\nu=\beta c \lambda$, where $\beta$ is a constant in the range of 1/3-1/2
depending on the situation, $c$ mean thermal speed of molecules. In DSMC, $\beta$ is not a constant but is
calculated automatically. Note that $\beta$ depends on the direction
considered in the case of rotating gas as in accretion discs. Note also the
similarity between the above formula to the famous alpha model in accretion
disc theory.

Viscous heating is included automatically in the calculation. Thermal
conduction is also embedded in the system. It may not be important in
two-dimensional simulations, but it can be used as a model of radiative
diffusion in three-dimensional cases. We do not consider the thermal
diffusion effect in the present work.

\section{Accretion disc in a close binary system}
We have performed many three-dimensional DSMC simulations of an accretion disc in a close binary system. Using an adiabatic equation of state results in the gas being unrealistically hot. With $\gamma=5/3$, this even prevents the formation of an accretion disc. In the present work, however, we include cooling. Although it is possible to consider a realistic cooling function,
we consider two simple models of cooling in the present work: inelastic
collision between particles or a drag to particle motion. 
In the first case, we
assume that the particle velocity is reduced, after a collision, by some constant
factor $e$. In the later case, we reduce the thermal speed of the molecules by a factor $f$ at each time step. Note that in either case the
cooling rate depends on the choice of the size of time step, so our
treatments here are rather phenomenological. In future work, we will consider
more realistic cooling function.

Because of lack of space, the results of our simulations will be presented elsewhere.

\end{document}